\documentclass[prl, showpacs, twocolumn]{revtex4}

\usepackage{amsmath}
\usepackage{amssymb}
\usepackage{graphicx}
\usepackage[caption=false]{subfig}

\newcommand{\pb}{\textbf{p} }
\newcommand{\rb}{\textbf{r} }
\newcommand{\nb}{\textbf{n} }

\begin{document}

\title{Relativistic electron Wigner crystal formation in a cavity for electron acceleration}
\author{Johannes Thomas$^1$}\email{thomas@tp1.uni-duesseldorf.de}
\author{Marc M. G\"unther$^{2,3}$}\email{m.guenther@gsi.de}
\author{Alexander Pukhov$^{1}$}

\affiliation{$^1$Institut f\"ur Theoretische Physik I, Heinrich-Heine-Universit\"at D\"usseldorf, D-40225  D\"usseldorf, Germany}
\affiliation{$^2$GSI Helmholtzzentrum f\"ur Schwerionenforschung GmbH, D-64291 Darmstadt, Germany}
\affiliation{$^3$Helmholtz-Institut Jena, D-07743 Jena, Germany}


\begin{abstract}
It is known that a gas of electrons in a uniform neutralizing background can crystallize and form a lattice if the electron density is less than a critical value. This crystallization may have two- or three-dimensional structure. Since the wake field potential in the highly-nonlinear-broken-wave regime (bubble regime) has the form of a cavity where the background electrons are evacuated from and only the positively charged ions remain, it is suited for crystallization of trapped and accelerated electron bunches. However, in this case, the crystal is moving relativistically and shows new three-dimensional structures that we call relativistic Wigner crystals. We analyze these structures using a relativistic Hamiltonian approach. We also check for stability and phase transitions of the relativistic Wigner crystals.
\end{abstract}

\maketitle

In the past, many acceleration concepts for electrons were investigated and developed. In the field of plasma-based electron acceleration we distinguish two principle methods: the particle beam-driven plasma wake field acceleration (PWFA) \cite{Rosenzweig1988,JoshiScAm2006,Hidding2012} and the laser-driven plasma wake field acceleration (LWFA). The latter can form a highly non-linear broken wave which leads to an electronic plasma cavity (``bubble'') \cite{Pukhov2002}. A similar wake field structure can be created in PWFA by a dense charged particle beam in the "blowout regime". These regimes could be used as effective electron accelerators with various  possible electron injection schemes into the plasma cavity. An overview of the state of the art laser plasma accelerators is given by V. Malka \cite{Malka2012}.

An interesting aspect of the bubble regime is that it forms a strong nearly harmonic potential (electric field strength of more than 100 GV/m) which focuses the injected electrons to the center of the cavity. This strongly focusing potential may lead to the formation of regular structures in the cloud of trapped electrons  known as electron Wigner crystals \cite{Wigner1934, Crandall1971, Meissner1976, Dubin1999}.  

The relativistic velocity of the bubble system will result in a relativistic electron crystal state. Non-relativistic Wigner crystals have been observed, e.g., in dusty plasmas \cite{Morfill2009}. We find comparable non-relativistic Wigner crystals in planes normal to the laser propagation direction while near the bubble origin the crystallization displays an essentially new structure. From the experimental point of view it will be a formidable challenge to detect the electron crystal formation. One of the first experimental approaches to investigate the electron bunch structure in the bubble regime has been described recently by M. Schnell et al. \cite{Schnell2012}. Another approach is a spectroscopy method based on Compton back-scattering, using a high resolution gamma-ray spectrometry \cite{Guenther2014}.

An important application for relativistic electron bunches is the realization of novel bright sources of short wavelength radiation. In the conventional Compton sources, a counterpropagating laser pulse is back-scattered from an unstructured relativistic electron bunch. This produces a spatially incoherent beam of back-scattered gamma-photons with a wide energy spectrum emitted into a rather broad solid angle \cite{Petrillo2012}. A regularly structured electron crystal  within the plasma cavity could greatly enhance the brightness and coherence of the Compton $\gamma-$source \cite{Guenther2014}.

In the following we discuss the  inner structure of an electron bunch injected into the bubble. Different from the widely used particle-in-cell (PIC) simulations, we  consider individual electrons as classical point-like particles with the physical charge and mass. This information is usually lost in the PIC method, where numerical macro-particles are used. Thus, we add the interaction between individual electrons to one model of bubble fields described by the recent bubble models \cite{Whittum1992, Kostyukov2004, Lu2006, Lu2007, Kostyukov2009, Kalmykov2009, Kostyukov2010, Yi2011, Kalmykov2011}. First we discuss equilibrium two dimensional Wigner crystal states on planes normal to laser propagation direction for cold electrons and then present new relativistic crystal states in a full 3D model. Finally we consider the impact of the finite bunch emittance (an analogue of transverse temperature) on stability of the crystalline electron structures and their phase transitions.

The most successful kinetic plasma simulation methods are the so called particle-in-cell (PIC) codes which use numerical macro-particles. The numerical macro-particles can be used to represent all physical particles that are found in the plasma (electrons, ions, and neutral atoms). They have the same charge-to-mass ratio as the real physical particles and move along the same trajectories in the mean electromagnetic fields. This makes PIC codes an efficient and valuable tool for plasma simulations, particularly for electron acceleration simulations in the bubble regime. However, PIC codes also have disadvantages because the number of the numerical macro particles is much smaller than that of the real plasma particles. Thus, a single macro-particle substitutes for many physical particles so that the information about the point-like nature of the physical particles is lost.  Also, the numerical noise pertinent to the PIC codes may increase the particle temperature artificially. Thus, it is difficult to resolve the inner structure of electron beams or their equilibrium distribution using the standard PIC methods. 

For our purposes, we choose a different, semi-analytical approach. A bunch of bubble models has been developed recently from which we chose the phenomenological bubble model first suggested in \cite{Kostyukov2004, Kostyukov2009} and extend it by electromagnetic interaction between the trapped particles. From a mathematical point of view we add to the known bubble Hamiltonian \cite{Kostyukov2010} (which we should call the free bubble Hamiltonian)  an interaction term between the electrons that is calculated from retarding potentials, see e.g.   \S65 in the textbook by Landau-Lifschiz \cite{Landau2009}. Thus, the Hamiltonian of the system of interacting charged particles $H$ splits up into two parts $H=H_0 + H_I$. $H_0$ is the Hamiltonian of non-interacting particles in the potential of the bubble. $H_I$ describes the interparticle interaction. 
\begin{figure}[t]
	\centering
		\subfloat[]{\label{fig:1940_1}\includegraphics[width=0.24\textwidth]{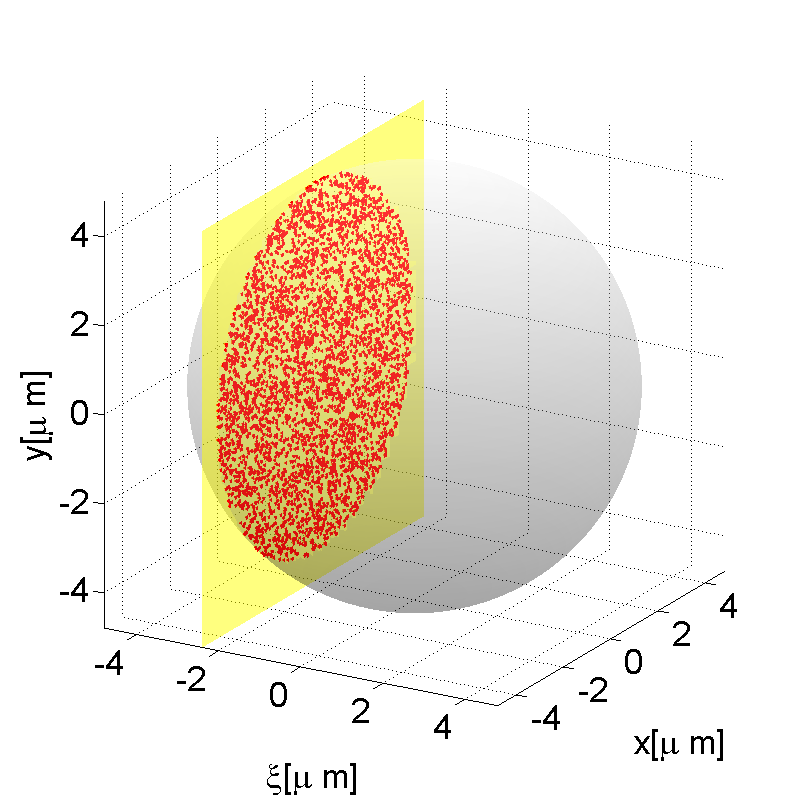}}
		\hfill
		\subfloat[]{\label{fig:1940_150}\includegraphics[width=0.24\textwidth]{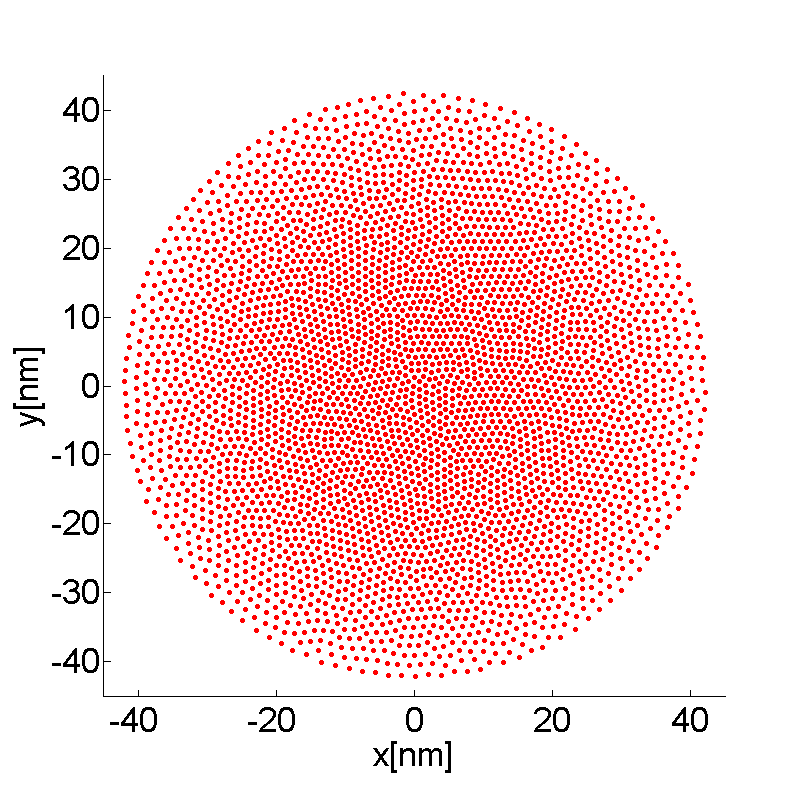}}\\
		\subfloat[]{\label{fig:1940_150_zoom}\includegraphics[width=0.24\textwidth]{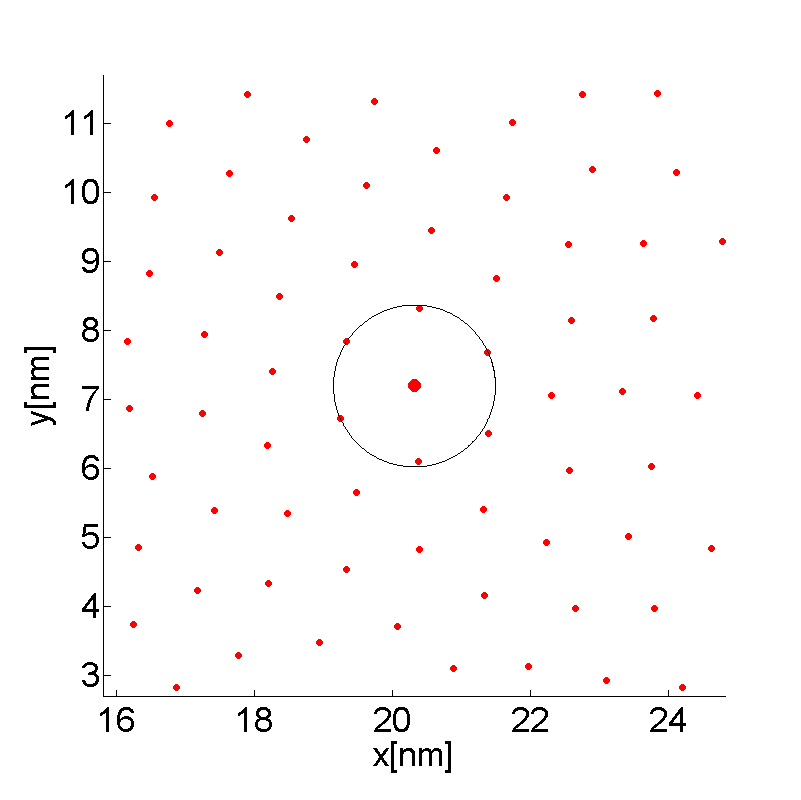}}
		\hfill
		\subfloat[]{\label{fig:n_fit}\includegraphics[width=0.24\textwidth]{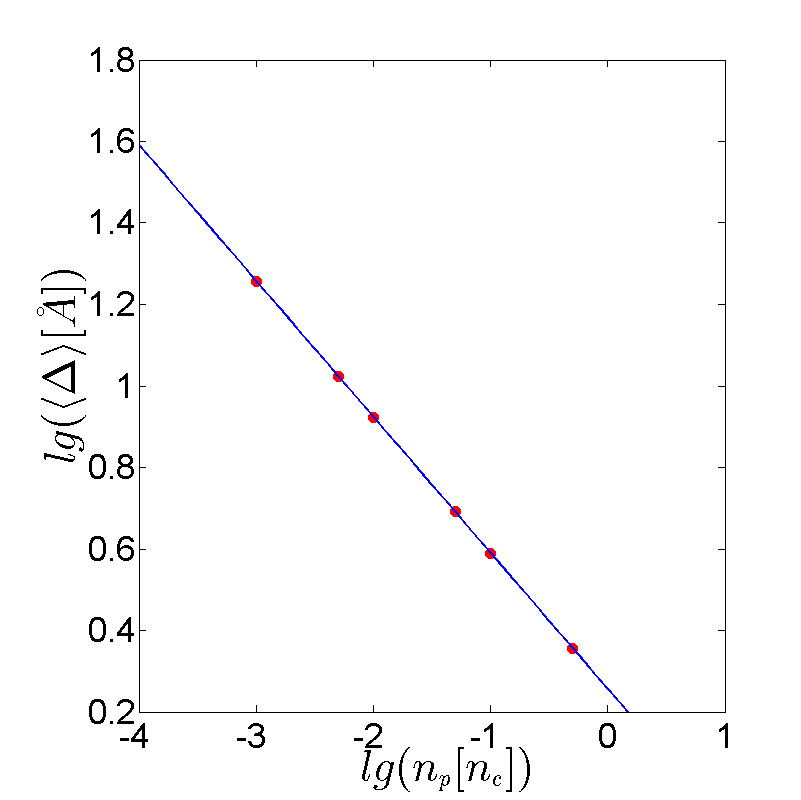}}
	\caption{\protect\subref{fig:1940_1} Initial electron distribution on a plane normal to laser propagation direction. \protect\subref{fig:1940_150} Front view on crystal that minimizes energy. \protect\subref{fig:1940_150_zoom} Inner structure of the crystal. \protect\subref{fig:n_fit} $\bar{\Delta}$ dependence on $n_p[n_c]$}
	\label{fig:1940}
\end{figure}

Throughout the paper, we use conventional relativistically normalized variables. The time is normalized with the plasma frequency $\omega_p$, the lengths are normalized with the plasma wavenumber  $k_p=\omega_p/c$, the momenta are normalized with $m_e c$, where $m_e$ is the electron mass.
The canonical transformation to the co-moving coordinate $\xi=z-Vt$ of the Hamiltonians $H_0$ and $H_I$ gives the total Hamiltonian

\small
\begin{multline}
H = \sum_i \left( \gamma_i + \frac{1}{2}\Phi_i - V\Pi_{x_i} \right) + \frac{r_e}{\lambda_{pe}}\sum_{i>j} \frac{1}{\Delta_{ij}} \\
-\frac{r_e}{\lambda_{pe}}\sum_{i>j} \frac{1}{2\Delta_{ij}}\left[\frac{\pb_i}{\gamma_i}\cdot\frac{\pb_j}{\gamma_j} + \left(\frac{\pb_i}{\gamma_i}\cdot\nb_{ij}\right) \left(\frac{\pb_j}{\gamma_j}\cdot\nb_{ij}\right)\right].	\label{eqn:HI}
\end{multline}
\normalsize
Here $\gamma_i$ is the electron relativistic $\gamma-$factor, $\mathbf{\Pi}_i=\pb_i-\dot{\Phi}_i/2$ is the canonical momentum, $\pb_i$ is the kinetic momentum, $r_e=e^2/(2\pi m_ec^2)$ is the classical electron radius, $\lambda_{pe}$ is the plasma wave length, $\Phi_i = (|\rb_i|^2-R(t)^2)/4$ is the bubble potential, and $\vec{\Delta}_{ij} = \rb_i - \rb_j$, $\nb_{ij} = \vec{\Delta}_{ij}/\Delta_{ij}$. According to Wigner, a spatial electron distribution that minimizes this Hamiltonian obeys a special symmetry. Therefore we shift the electron positions in the bubble until a local minimum is reached. The algorithm we use for the shift is the steepest descent method
\small
\begin{align}
x^{k+1} = x^k - \nabla_{x^k} H\cdot \Delta t	\label{eqn:min3}
\end{align}
\normalsize
with $x^k = (\rb_1^k,\cdots,\rb_N^k)$ as the three dimensional spatial distribution after the $k$-th iteration. $\Delta t$ is an appropriately chosen iteration step.
Next we iterate this algorithm for a 2D distribution of 4000 cold electrons ($\pb_\bot=0$) with fixed $\xi$-coordinate on a plane normal to laser propagation direction (highlighted area in Fig.\ref{fig:1940_1}). The resulting spatial configuration is known to be the equilibrium distribution of electrons. 

We assume that the driving laser pulse that generated the bubble has wavelength $\lambda_l=800$~nm and focal spot size radius $R=6\lambda_l$. The plasma density $n_p$ is $0.5\%$ of critical density $n_c=2\cdot10^{21}/$cm$^3$. After a local minimum of (\ref{eqn:HI}) is found (see Fig.\ref{fig:1940_150}) the electrons are arranged in a structure that has already been identified as Wigner crystals in fields of non-relativistic dusty plasmas \cite{Melzer1996, Melzer2003, Morfill2009}. The diameter of the distribution is $800$~nm which means that it occupies a $10^4$ times smaller area than the bubble cross section. The mean areal density of the distribution is $7\cdot10^{13}\text{ cm}^{-2}$, which is close to $n_c$. A closer zoom into the distribution in Fig.\ref{fig:1940_150_zoom} reveals that most electrons lay in a hexagonal structure with a mean electron distance $\bar{\Delta}\approx 10$~\AA\, (see black circle around the large electron).

For future experiments this structure is extremely interesting because a highly ordered bunch structure is related to a high spatial coherence. Thus it is crucial to know the scaling of the mean electron distances with laser and plasma parameters. To find the scaling, we calculate the mean distance in a parameter scan over plasma density $n_p$ and electron number $N$. An exemplary part of the scan for $\lambda_{l}=800$~nm and $N=4000$ is shown as red dots in Fig.\ref{fig:n_fit} together with a fit as the blue line.

The dependencies we found can be summarized as
\small
\begin{align}
\bar{\Delta}[\text{\AA}]=6.9n_{p}^{-1/3}N^{-0.15}. \label{eqn:scaling}
\end{align}
\normalsize

\begin{figure}[t]
	\centering
		\subfloat[]{\label{fig:3326_1}\includegraphics[width=0.24\textwidth]{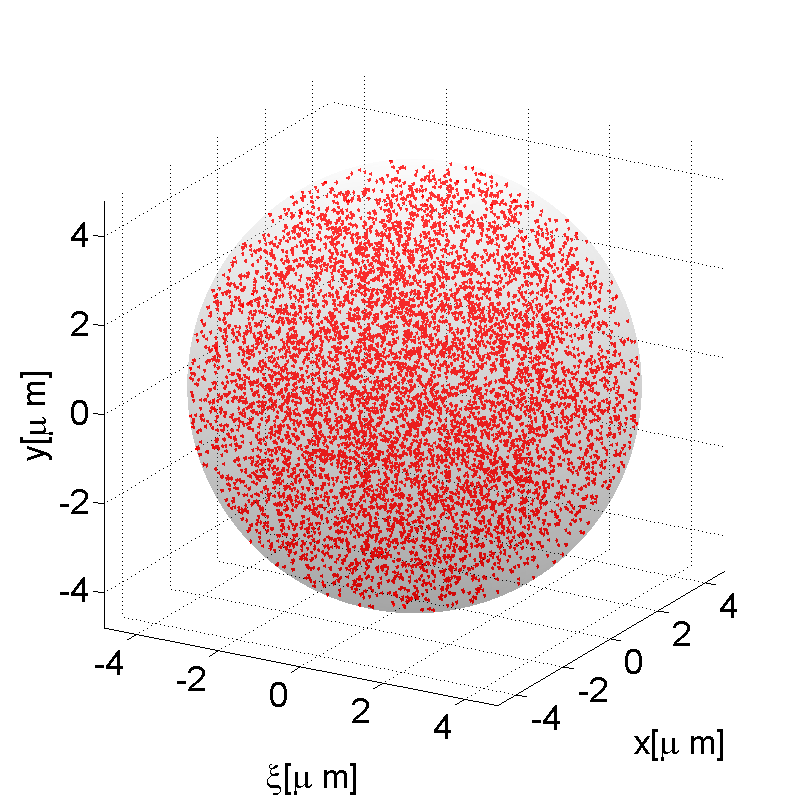}}
		\hfill
		\subfloat[]{\label{fig:3326_150}\includegraphics[width=0.24\textwidth]{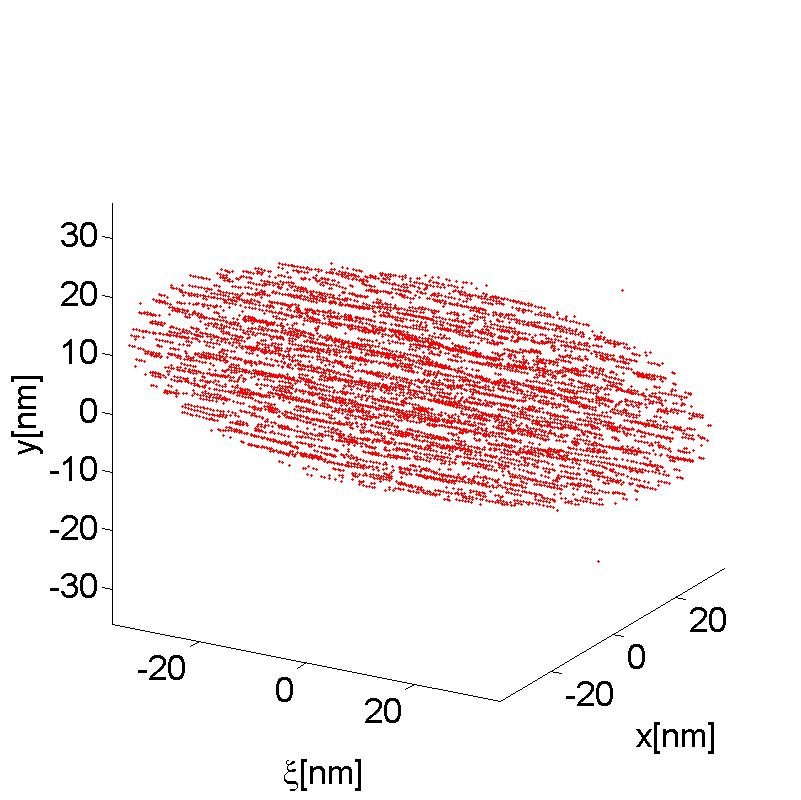}}\\
		\subfloat[]{\label{fig:3326_k}\includegraphics[width=0.24\textwidth]{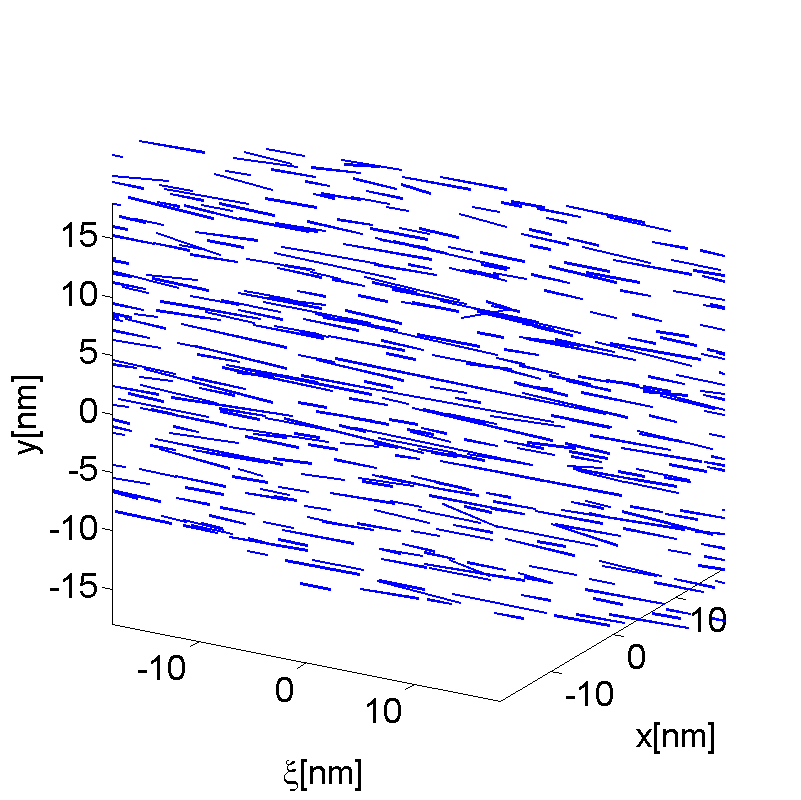}}
		\hfill
		\subfloat[]{\label{fig:3326_pk}\includegraphics[width=0.24\textwidth]{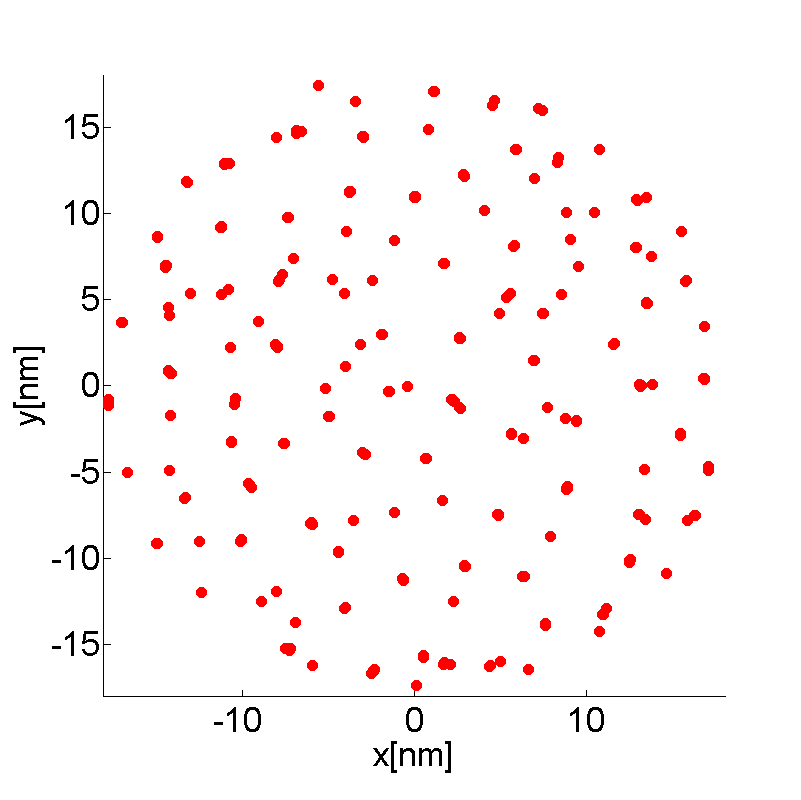}}
	\caption{\protect\subref{fig:3326_1} Electron distribution in the bubble volume. \protect\subref{fig:3326_150} Distribution that minimizes energy. \protect\subref{fig:3326_k} Identified filaments as inner structure of the distribution. \protect\subref{fig:3326_pk} Projection of electrons on the $x$-$y$-plane.}
	\label{fig:3326}
\end{figure}

To expand the discussion of Wigner crystals in relativistic structures on electron distributions in the whole bubble volume, we prepare a random electron distribution like the one shown in Fig.\ref{fig:3326_1} and shift the electron positions till the Hamiltonian (\ref{eqn:HI}) is minimized. Similar to the 2D case, we use the steepest descent method (\ref{eqn:min3}) with the difference that now the electron positions in all three dimensions are varied. As we know, the resulting spatial structure should exhibit a symmetric structure. What we find for 8000 cold electrons in a bubble generated by a $\lambda_l=800$~nm, $R=6\lambda_l$ laser pulse in a $n_p=0.5\%n_c$ plasma is shown in Fig.\ref{fig:3326_150}. The electrons are focused near the bubble origin in a volume that is $10^6$ times smaller than the bubble. The inner structure consists of filaments oriented parallel to the $\xi$-axis around the origin.

To analyze the filaments automatically we define them as follows:
Let $\Delta_{con}$ (\textit{connection length}) be an arbitrary chosen positive number. Then two electrons are \textit{connected}, if the distance between them is smaller or equal $\Delta_{con}$ and if there is no 3rd electron that is closer to both. The \textit{starting point of a filament} is the electron in a filament, that has no left connected neighbor. An \textit{end point of a filament} is the electron in a filament that has no right connected neighbor. A filament in turn is the set of connected electrons with exactly one starting point and exactly one end point. The \textit{length of a filament} is the distance between its starting point and its end point.

A figure that illustrates how this definition fits to an actual electron distribution is shown in Fig.\ref{fig:3326_k}. Here blue lines link start and end points of found filaments. The minimal filament length in every 3D crystal is less than 1\AA. The maximum filament length is in a range of 20~nm. 

A possible reason why long electron filaments minimize the energy of the system is the deformation of the electromagnetic field of the rapidly moving electrons. According to a proper Lorentz transformation from the electrons rest frame into the laboratory frame, the electromagnetic field of the accelerated electrons greatly reduces in $\xi$-direction. This leads to a situation in which the electromagnetic coupling of the electrons is effectively restricted to a transverse interaction.

Another aspect of the 3D symmetry we found can be seen in the projection of electron positions with $|\xi|\leq 1$~nm on the $x$-$y$-plane in Fig.\ref{fig:3326_pk}. Here the electrons lie on circles around the origin. A similar profile is known from non relativistic dusty plasmas for three dimensional cluster balls \cite{Arp2004, Arp2005}. For us this is important because the 2D structures discussed above obey the same arrangement at the periphery and prove that the filamentation in propagation direction is a direct and measurable relativistic effect.

\begin{figure}[t]
	\centering
		\subfloat[]{\label{fig:1941_dyn}\includegraphics[width=0.24\textwidth]{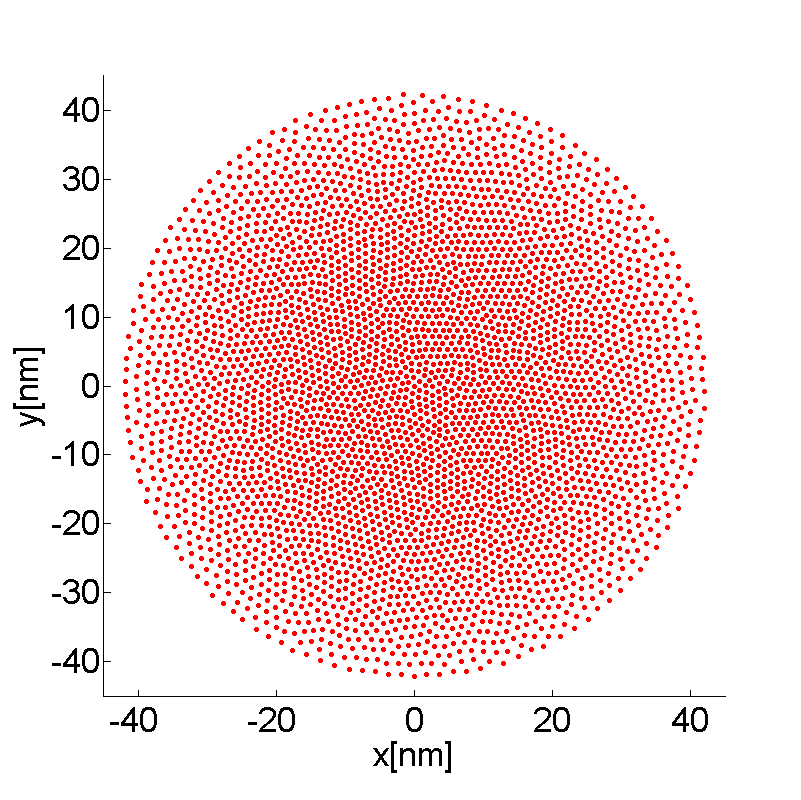}}
		\hfill
		\subfloat[]{\label{fig:1950_dyn}\includegraphics[width=0.24\textwidth]{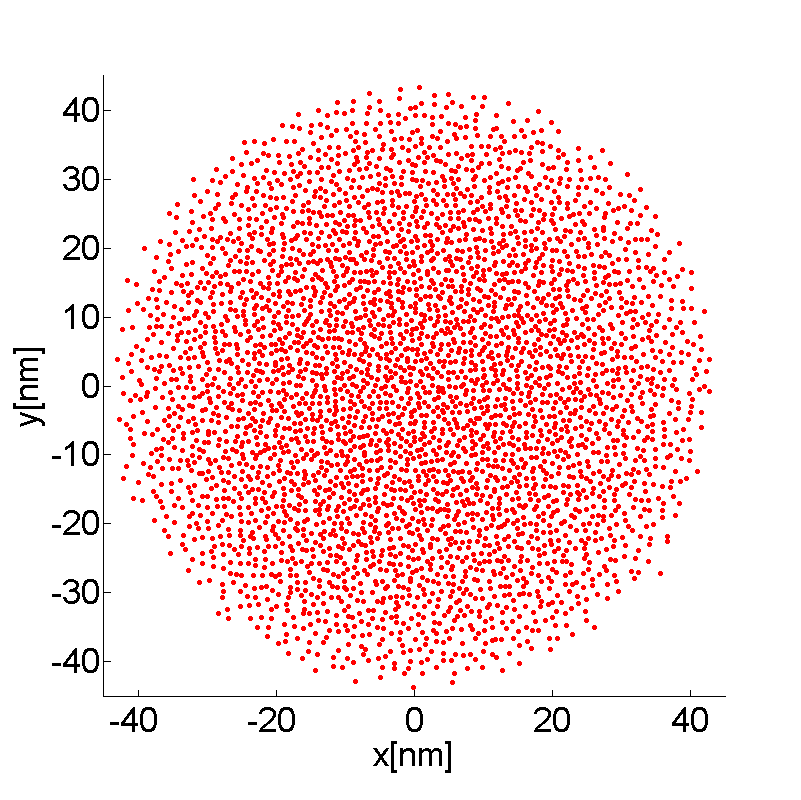}}
	\caption{Kinetic simulation with same parameters as in Fig.\ref{fig:1940_150} after a simulation time of $t=25\omega_p$ for \protect\subref{fig:1941_dyn} $\epsilon=0.05$~mm\,mrad and \protect\subref{fig:1950_dyn} $\epsilon=0.5$~mm\,mrad.}
	\label{fig:2D_dyn}
\end{figure}
Another important feature we discuss for relativistic Wigner crystals in the bubble regime is the influence of a finite emittance to the stability of the found structures or, in other words, the finite transverse temperature that defines the transition from solid crystals to a liquid phase. Mathematically we solve the equations of motion for electrons that are initially in a crystal structure, generated by algorithm (\ref{eqn:min3}). The equations of motion calculated from (\ref{eqn:HI}) are
\small
\begin{align}
\frac{d\pb_i}{dt} =& \frac{1}{2}\frac{d\Phi_i}{dt}\mathbf{e}_x - \frac{\partial H}{\partial \rb_i}, & \frac{d\rb_i}{dt} =& \frac{\partial H}{\partial \mathbf{\Pi_i}}.
\end{align}
\normalsize
If we solve these equations for electrons that were fixed on a plane initially (cmp. the 2D case), we distinguish between two cases. First, if the emittance of the electron bunch is small enough, the 2D Wigner Crystal is accelerated in positive $\xi$-direction while the crystalline structure is preserved. The center of mass of the ensemble gets accelerated like a single electron on the $\xi$-axis, except that no electron performs betatron oscillations. For a momentum ratio of $\sqrt{\langle\pb^2_\bot\rangle}=0.001\%|\pb|$ and an acceleration time of $t=25\omega_p$ Fig.\ref{fig:1941_dyn} proves that indeed the initial configuration for the same laser plasma parameters as in the 2D case, is a stable configuration. If one takes into account that trapped electron bunches in the bubble can contain charges in the nC range [19] and not just 4000 particles we were able to simulate, the corresponding normalized beam emittance is $\epsilon=0.05$~mm\,mrad. For a ten times larger beam emittance and an acceleration time of again $t=25\omega_p$ Fig.\ref{fig:1950_dyn} shows that the crystal state is destroyed without that the ensemble runs out its confining border. This state we call a 2D electron liquid in the bubble regime. Since electron beams with a normalized emittance in the order of $0.1 - 1$~mm\,mrad can be produced \cite{Weingartner2012} an experimental verification of both liquid and crystal electron states in the bubble regime should be feasible.

\begin{figure}[t]
	\centering
		\subfloat[]{\label{fig:3327_d}\includegraphics[width=0.24\textwidth]{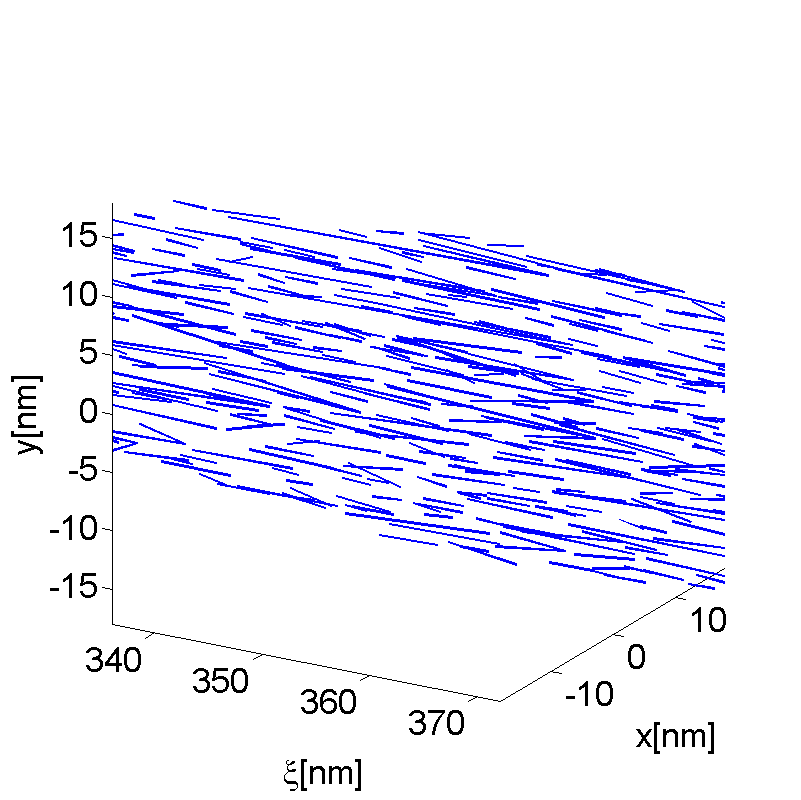}}
		\hfill
		\subfloat[]{\label{fig:3336_d}\includegraphics[width=0.24\textwidth]{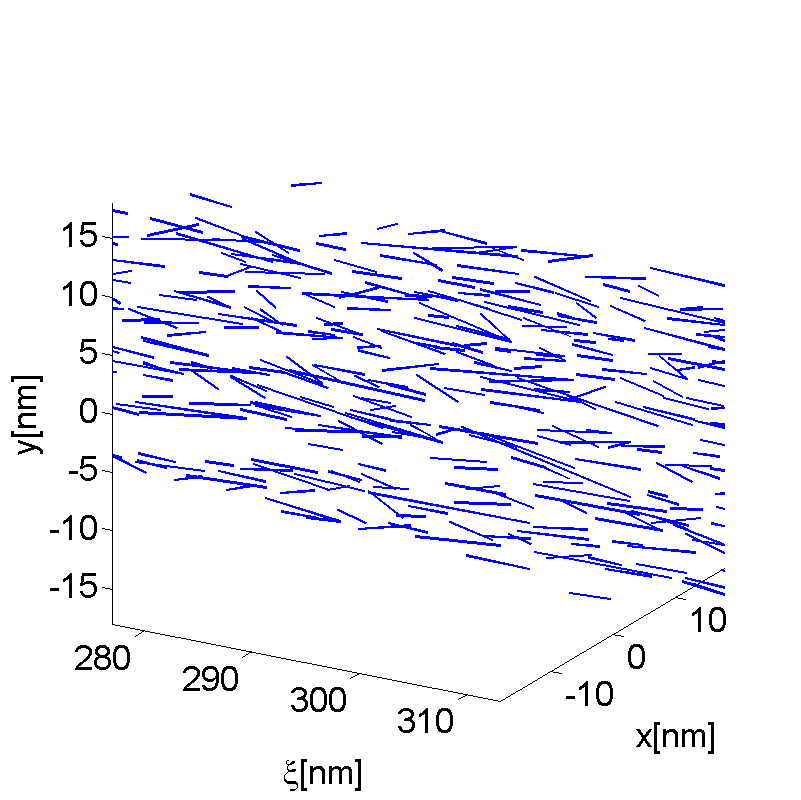}}\\
		\subfloat[]{\label{fig:3327_pd}\includegraphics[width=0.24\textwidth]{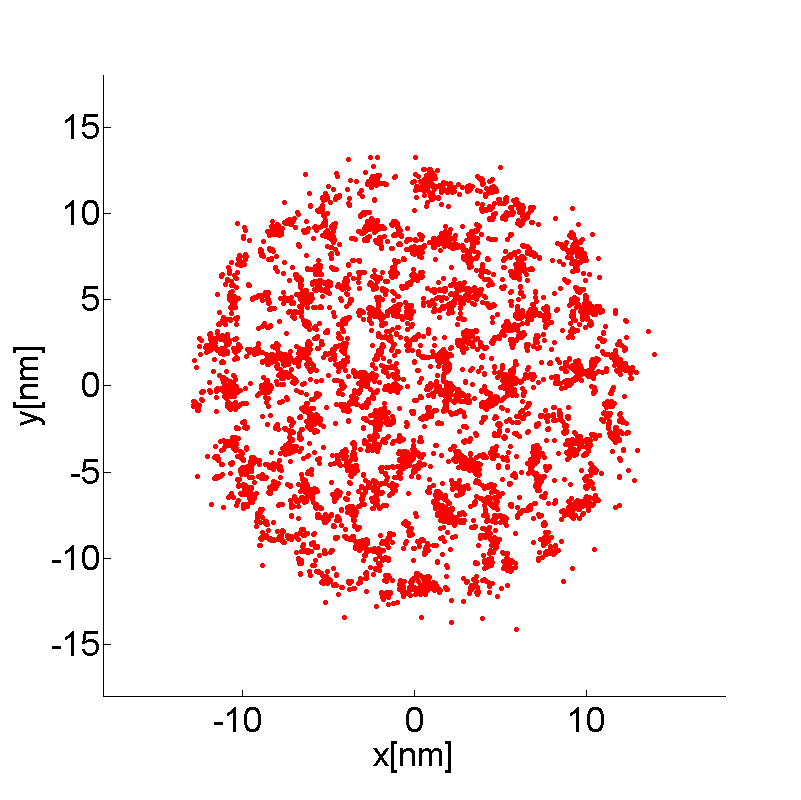}}
		\hfill
		\subfloat[]{\label{fig:3336_pd}\includegraphics[width=0.24\textwidth]{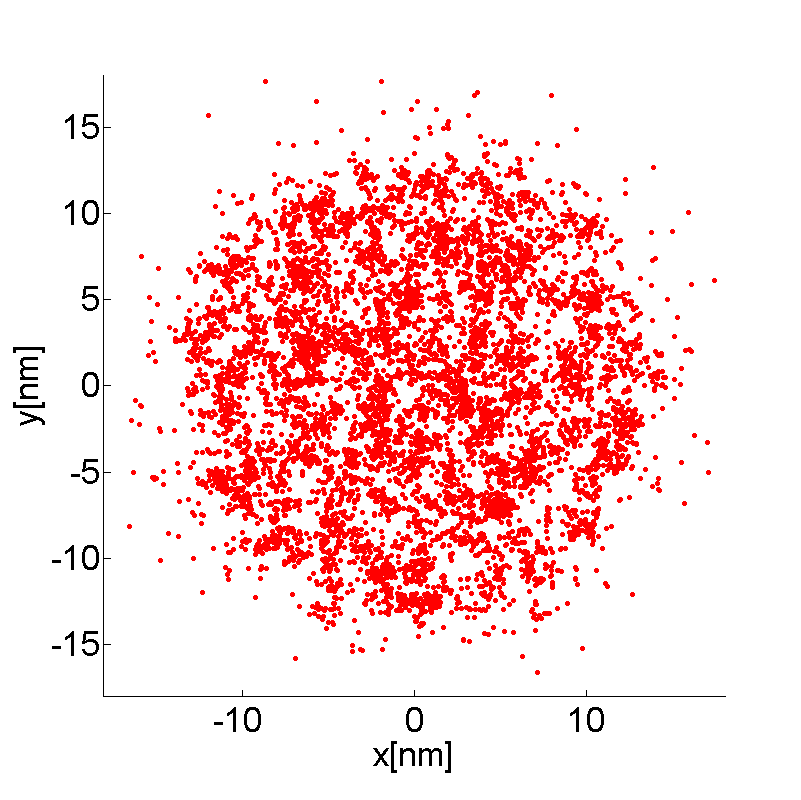}}\\		
	\caption{\protect\subref{fig:3327_d},\protect\subref{fig:3327_pd} Kinetic simulation with same parameters as in Fig.\ref{fig:3326_k} after a simulation time of $t=60\omega_p$ for $\epsilon=0.03$~mm\,mrad and projection of all electrons to the $x$-$y$-plane. \protect\subref{fig:3336_d},\protect\subref{fig:3336_pd} Kinetic simulation with same parameters as in Fig.\ref{fig:3326_k} after a simulation time of $t=50\omega_p$ for $\epsilon=0.3$~mm\,mrad and projection of all electrons to the $x$-$y$-plane.}
	\label{fig:3D_dyn}
\end{figure}

Now we solve the equations of motion for electrons that are initially in a 3D relativistic Wigner crystal state. Again, we distinguish between the two cases $\sqrt{\langle\pb^2_\bot\rangle}=10^{-5}|\pb|$ corresponding to $\epsilon=0.03$~mm\,mrad and $\sqrt{\langle\pb^2_\bot\rangle}=10^{-4}|\pb|$ corresponding to $\epsilon=0.3$~mm\,mrad. For the lower normalized beam emittance and a simulation time of $t=60\omega_p$ the elongation of the ensemble increases and the filaments get closer and longer (see Fig.\ref{fig:3327_d}). Fig.\ref{fig:3327_pd} is a projection of all electrons onto the $x$-$y$-plane and reveals that the circular structure in normal direction is also preserved. Contrary to that we find for the higher beam emittance and a simulation time of $t=50\omega_p$ shorter and less parallel filaments in Fig.\ref{fig:3336_d}. The symmetry in $\xi$-direction thus dissolves but the occupied volume is the same. The projection in Fig.\ref{fig:3336_pd} shows a similar melting. Altogether we call the electron configuration in these two pictures a 3D relativistic electron fluid with defined boundaries and without a symmetric inner structure.

Summarizing we point out that relativistic effects on Wigner crystals in the bubble appear only along the propagation direction. In the normal direction we see non relativistic crystals both in 2D and 3D configurations. A preliminary discussion of the impact of a non-zero bunch emittance on crystal states yields that also relativistic Wigner crystals may be realized only for electron temperatures smaller than a certain critical temperature. An analysis of this special temperature at which a relativistic electron crystal melts and becomes an electron fluid will be done in future articles.

The external potential we apply is referred to the bubble regime. Nevertheless other relativistic harmonic potentials and especially those from the blow out regime are equally suited to describe relativistic Wigner crystals.

This work has been supported by the Deutsche Forschungsgemeinschaft via GRK 1203 and SFB TR 18. This work has been supported by EU EuCARD-2 and ANAC2 -projects.

\bibliographystyle{prsty}
\bibliography{Kristalle}

\end{document}